\documentclass[8.5pt,twoside,twocolumn]{article}
\usepackage[super,sort&compress,comma]{natbib}
\usepackage{mhchem}
\usepackage{mathrsfs}
\usepackage{amsmath,amssymb,latexsym}
\usepackage{bm}
\usepackage{times,mathptmx}
\usepackage{sectsty}
\usepackage{balance}

\usepackage{graphicx} %eps figures can be used instead
\usepackage{lastpage}
\usepackage[format=plain,justification=raggedright,singlelinecheck=false,font=small,labelfont=bf,labelsep=space]{caption}
\usepackage{fancyhdr}

\begin{document}

\thispagestyle{plain}
\fancypagestyle{plain}{
\fancyhead[L]{\includegraphics[height=8pt]{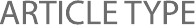}}
\fancyhead[C]{\hspace{-1cm}\includegraphics[height=20pt]{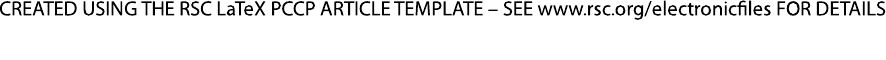}}
\fancyhead[R]{\includegraphics[height=10pt]{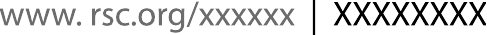}\vspace{-0.2cm}}
\renewcommand{\headrulewidth}{1pt}}
\renewcommand{\thefootnote}{\fnsymbol{footnote}}
\renewcommand\footnoterule{\vspace*{1pt}%
\hrule width 3.4in height 0.4pt \vspace*{5pt}}

\newcommand{\um}{~\mu\mathrm{m}}
\newcommand{\erf}{\mathrm{Erf}}
\newcommand{\re}{\mathrm{Re}}
\newcommand{\sinc}{\mathrm{sinc}}
\setcounter{secnumdepth}{5}

\makeatletter
\def\subsubsection{\@startsection{subsubsection}{3}{10pt}{-1.25ex plus -1ex minus -.1ex}{0ex plus 0ex}{\normalsize\bf}}
\def\paragraph{\@startsection{paragraph}{4}{10pt}{-1.25ex plus -1ex minus -.1ex}{0ex plus 0ex}{\normalsize\textit}}
\renewcommand\@biblabel[1]{#1}
\renewcommand\@makefntext[1]%
{\noindent\makebox[0pt][r]{\@thefnmark\,}#1}
\makeatother
\renewcommand{\figurename}{\small{Fig.}~}
\sectionfont{\large}
\subsectionfont{\normalsize}

\fancyfoot{}
\fancyfoot[LO,RE]{\vspace{-7pt}\includegraphics[height=9pt]{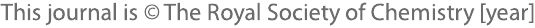}}
\fancyfoot[CO]{\vspace{-7.2pt}\hspace{12.2cm}\includegraphics{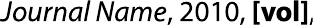}}
\fancyfoot[CE]{\vspace{-7.5pt}\hspace{-13.5cm}\includegraphics{RF}}
\fancyfoot[RO]{\footnotesize{\sffamily{1--\pageref{LastPage} ~\textbar  \hspace{2pt}\thepage}}}
\fancyfoot[LE]{\footnotesize{\sffamily{\thepage~\textbar\hspace{3.45cm} 1--\pageref{LastPage}}}}
\fancyhead{}
\renewcommand{\headrulewidth}{1pt}
\renewcommand{\footrulewidth}{1pt}
\setlength{\arrayrulewidth}{1pt}
\setlength{\columnsep}{6.5mm}
\setlength\bibsep{1pt}

\twocolumn[
  \begin{@twocolumnfalse}
\noindent\LARGE{\textbf{Probing shear-induced rearrangements in Fourier Space. I. Dynamic Light Scattering
}}
\vspace{0.6cm}

\noindent\large{\textbf{S. Aime,$^{\ast}$\textit{$^{a}$} and L. Cipelletti\textit{$^{a}$  }
}}\vspace{0.5cm}
%Please note that \ast indicates the corresponding author(s) but no footnote text is required.

%\noindent\textit{\small{\textbf{
%Last update: \today}}}

\vspace{0.6cm}
%Please do not change this text.

\noindent \normalsize{Understanding the microscopic origin of the rheological behavior of soft matter is a long-lasting endeavour. While early efforts concentrated mainly on the relationship between rheology and structure, current research focuses on the role of microscopic dynamics. We present in two companion papers a thorough discussion of how Fourier space-based methods may be coupled to rheology to shed light on the relationship between the microscopic dynamics and the mechanical response of soft systems. In this first companion paper, we report a theoretical, numerical and experimental investigation of dynamic light scattering coupled to rheology. While in ideal solids and simple viscous fluids the displacement field under a shear deformation is purely affine, additional non-affine displacements arise in many situations of great interest, for example in elastically heterogeneous materials or due to plastic rearrangements. We show how affine and non-affine displacements can be separately resolved by dynamic light scattering, and discuss in detail the effect of several non-idealities in typical experiments.}
\vspace{0.5cm}
\end{@twocolumnfalse}
]

%Footnotes
%\footnotetext{\dag~Electronic Supplementary Information (ESI) available: [details of any supplementary information available should be included here]. See DOI: 10.1039/b000000x/}

%Please use \dag to cite the ESI in the main text of the article.
%If you article does not have ESI please remove the the \dag symbol from the title and the above footnotetext.

\footnotetext{\textit{$^{a}$ L2C, University of Montpellier, CNRS, Montpellier, France. E-mail: stefano.aime@umontpellier.fr}}

%additional addresses can be cited as above using the lower-case letters, c, d, e... If all authors are from the same address, no letter is required

\section{Introduction}
\label{sec:introduction}

	Soft matter systems are significantly deformed or even flown by applying modest forces, corresponding to stresses comparable to those due to thermal fluctuations. Consequently, the rheological behavior of soft systems is of great fundamental interest and of paramount importance in technological applications. This makes soft matter rheology the object of a sustained research effort, both in academia and in industry~\cite{larson_structure_1998}. One of the key questions actively investigated concerns the interplay between the rheological properties of soft systems and their microscopic structure and dynamics. Indeed, structure and dynamics determine the microscopic relaxation processes responsible for the macroscopic mechanical properties of a system and are in turn profoundly modified by externally imposed stresses, e.g. in shear alignment or in shear thinning and thickening.

	The interplay between rheology and the microscopic structure and dynamics is currently studied intensively by numerical simulations, as well as experimentally. Separate experiments probing rheological properties on one hand and the microscopic structure and dynamics on the other hand are certainly informative and do provide valuable insight, especially in the linear regime and for stationary, non-thixotropic samples. However, simultaneous mechanical and microscopic measurements are highly desirable, in particular when probing the non-linear regime, where both the rheological response and the microscopic behavior are typically non-stationary and often exhibit sample-to-sample or even run-to-run strong fluctuations, e.g. in the onset of shear bands~\cite{divoux_transient_2010} or in plastic rearrangements with complex spatio-temporal patterns~\cite{knowlton_microscopic_2014,ghosh_direct_2017}. Optical and confocal microscopy are now used by several groups to investigate the sample evolution at the single particle scale, in conjunction with rheology, either by coupling a (confocal) microscope to a commercial rheometer~\cite{koumakis_yielding_2012,sentjabrskaja_creep_2015} or by using a dedicated shear cell~\cite{derks_confocal_2004,besseling_three-dimensional_2007,schall_structural_2007,lin_multi-axis_2014}.
	Microscopy methods allow the tracking of individual particles, thus providing one with the most complete microscopic information. However, microscopy comes with some limitations: typically, particle-based samples are required, and the particle size has to be larger than about $0.5~\um$; particles must be fluorescently labelled for confocal microscopy and tracking their position to better than about $0.1~\um$ requires special care and intensive image processing~\cite{bierbaum_light_2017}. Even more importantly, a high spatial resolution, a large field of view, and a high acquisition rate are conflicting requirements, so that one cannot resolve minute displacements with good time resolution while imaging a large portion of the sample. Fourier space techniques, and in particular scattering methods, are an attractive alternative to real-space measurements: although no information can be obtained on individual particles, scattering techniques allow a large sample volume to be probed, with no particular constraint on the temporal resolution. Moreover, particles in a wide range of sizes, from a few nm up to several microns, can be easily studied, as well as polymer- or surfactant-based systems. In this paper, we shall focus on scattering methods. In a companion paper~\cite{companion2} we will discuss Differential Dynamic Microscopy~\cite{cerbinoPRL,Giavazzi2014} (DDM) coupled to rheology. While DDM is based on microscopy, the data analysis is performed in Fourier space, using a formalism close to that of scattering techniques.

Historically, scattering methods coupled to rheology have been first used to investigate the sample structure~\cite{tolstoguzov_deformation_1974,hashimoto_apparatus_1986,van_egmond_time-dependent_1992}, but since the Nineties of the last century, dynamic scattering methods have become increasingly popular as a powerful tool to probe the microscopic dynamics of mechanically driven soft systems~\cite{wu_diffusing-wave_1990,gopal_nonlinear_1995,hebraud_yielding_1997}. Most experiments have been performed in the highly multiple scattering limit for visible light (Diffusing Wave Spectroscopy, DWS~\cite{weitz_diffusing-wave_1993}), thanks to the simplicity of the required optical layout and the high sensitivity of the method, which can detect motion on length scales as small as a fraction of a nm. A few experiments have been performed in the single scattering limit, typically (but not exclusively~\cite{ali_rheospeckle:_2016}) in a small-angle configuration where several scattering angles can be probed at the same time, using both visible light~\cite{tamborini_plasticity_2014} (Dynamic Light Scattering, DLS~\cite{berne_dynamic_1976}) and coherent X-ray radiation~\cite{leheny_rheo-xpcs_2015} (X-photon correlation spectroscopy, XPCS~\cite{madsen_beyond_2010}).

Most scattering experiments have been performed simultaneously to oscillatory rheology, using the so-called echo method~\cite{hebraud_yielding_1997,hohler_periodic_1997,petekidis_rearrangements_2002,leheny_rheo-xpcs_2015}, where the evolution of the sample microscopic configuration is measured stroboscopically, by comparing the pattern of the scattered intensity at each cycle, e.g. when the deformation is zero. This protocol allows the irreversible rearrangements to be highlighted and avoids the complications arising when affine and non-affine displacements are to be separately quantified. To address this key point, let us consider the geometry that we shall discuss in this paper, e.g. a plane-plane cell imposing a linear shear deformation, with the unit vectors $\left(\hat{u}_x, \hat{u}_y, \hat{u}_z\right)$ pointing in the shear velocity, shear vorticity and shear gradient directions, respectively. For both an ideal solid and a purely viscous fluid, the displacement field is affine:
%+++++++++++++++++++++++++++++++++++++++++++++++
\begin{equation}
    \Delta\bm{r}(\bm{r},\gamma)=\Delta\bm{r}^{\mathrm{(aff)}}(\bm{r}, \gamma) = (\bm{r} \cdot \hat{u}_z) \gamma \hat{u}_x \,,
    \label{eqn:AffineDeformation}
\end{equation}
%+++++++++++++++++++++++++++++++++++++++++++++++
where $e$ is the cell gap and $\gamma = \delta/e$ the shear deformation imposed by displacing the $z=e$ plane by an amount $\delta$ in the $\hat{u}_x$ direction. Non-affine displacements may arise as a consequence of heterogeneous elastic response~\cite{didonna_nonaffine_2005,basu_nonaffine_2011}, or as a result of non-linearities~\cite{liu_visualizing_2007}, or due to plastic rearrangements~\cite{hohler_periodic_1997}. They represent additional motion, typically not restricted to the direction of the imposed deformation, on top of the affine deformation field. %They typically involve additional motion on top of the affine deformation field; these additional displacements are not restricted to the direction of the imposed deformation.
Clearly, being able to discriminate between affine and non-affine dynamics is mandatory in order to fully characterize the complex interplay between rheology and microscopic dynamics, beyond the idealized case of Eq.~\ref{eqn:AffineDeformation}.

In DWS, the microscopic dynamics is quantified by measuring the change of phase of photons undergoing many scattering events in the sample. Since photons propagate following an isotropic, random walk-like path~\cite{weitz_diffusing-wave_1993}, they probe microscopic displacements along all directions. Hence, DWS is sensitive both to affine and non-affine displacements. For a given macroscopic deformation, it is possible to calculate theoretically the contribution of affine motion to the DWS signal~\cite{bicout_diffusing_1993,erpelding_diffusive_2008}. Any deviations from this behavior can then be ascribed to non-affine displacements. In practice, however, this approach is difficult to implement, because the approximations required to perform the calculation are not fully met in experiments~\cite{erpelding_diffusive_2008,nagazi_space-resolved_2017}, and because non-affine motion is typically small in comparison to the affine component. Diffusing Wave Spectroscopy is thus generally restricted to the case where the affine displacements vanish, as in the echo protocol or in stress relaxation tests~\cite{nagazi_cartographie_2017}, where a sample under load is kept at a fixed deformation.

Single scattering, by contrast, probes the dynamics along a well-defined direction, controlled by the experimental geometry. It is therefore possible to selectively probe motion along any of the $\left(\hat{u}_x, \hat{u}_y, \hat{u}_z\right)$ directions, thereby allowing affine and non-affine dynamics to be discriminated. This paves the way to measurements of the microscopic dynamics in a wide variety of rheological tests, including steady shear rate or creep tests, currently extensively used to understand the yield transition of amorphous soft solids~\cite{bonn_yield_2015,buchanan_matter_2017}, or oscillatory tests beyond the echo protocol, where surprising effects such as `reversible plasticity' may be unveiled by inspecting the dynamics throughout the whole oscillation cycle~\cite{keim_mechanical_2014,fiocco_encoding_2014}. As we shall show it in the following, the ability of single scattering to resolve motion along one single direction (e.g. along the vorticity direction $\hat{u}_y$) relies on several assumptions: the sample must be illuminated by an infinitely extended plane wave, the collection optics must be aberration-free, and the size of the detector along the direction orthogonal to the one selected (e.g. along the shear direction $\hat{u}_x$) must be vanishingly small. Clearly, these conditions cannot be strictly met in experiments.

In this paper, we derive analytical expressions that quantify the decay of correlation functions measured in DLS as a result of both affine and non-affine displacements, taking into account the effect of non-ideal experimental conditions. We provide guidelines for mitigating or correcting for the contributions to the correlation functions arising from non-ideal conditions and successfully test our predictions against experiments and numerical simulations. Although we will focus on simple shear coupled to a small-angle DLS setup, the results presented here are general and can be easily adapted to different experimental layouts, e.g. to a backscattering geometry. The rest of the paper is organized as follows: in Sec.~\ref{sec:MaterialsMethods} we briefly introduce the samples used for the tests, the DLS apparatus, the shear cell, and the simulation methods. In Sec.~\ref{sec:DLSresults} we develop a theoretical model of DLS under shear, starting from a discussion of the effect of a simple translation of the sample, and then presenting results for a purely affine deformation and the general case with both affine and non-affine microscopic dynamics. At each step, we show experimental and numerical data that validate our theoretical approach. We recapitulate our main findings and make some concluding remarks in Sec.~\ref{sec:conclusions}. For the reader's convenience, Appendix 1 lists all the symbols used in the paper, whereas Appendix 2 contains a more detailed derivation of the analytical results presented in the text.

\section{Materials and methods}
\label{sec:MaterialsMethods}
\subsection{Samples}

For studying the effect of a pure sample translation, quasi-2D random scatterers were prepared by sandblasting microscope glass slides (sandblaster Otelo OTMT with aluminium oxide particles of diameter $30 \um$). A commercial linear stage (Linear Stage UMR8.25A, by Newport) coupled to a stepper motor (Newport Precision Motorized Actuator LTA-HS) with a nominal precision of $1 \um$ was used to impose a controlled displacement in the $\hat{u}_x$ direction. The actual displacement was measured to within an accuracy of about 60 nm using the speckle imaging technique described in \cite{cipelletti_simultaneous_2013,aime_stress-controlled_2016}.

As a model sample for investigating 3D shear, we prepared polyacrylamide (PA) gels by polymerizing acrylamide monomers and (bis)acrylamide cross-linkers, using a free-radical polymerization reaction as described in~\cite{menter_p._acrylamide_2000}. $\mathrm{TiO}_2$ nanoparticles were added to the monomer solution at a concentration of around $0.01\%$. The particle diameter is $0.3 \um$ (resp., $0.5 \um$) for the scattering experiments (resp., for microscopy observations, see below). %\textbf{XXXStefano: ok particle size? XXX}.
In order to thoroughly disperse the $\mathrm{TiO}_2$ particles, the solution was sonicated for roughly 2h and then filtered before adding the initiators.

\subsection{Shear cell}

For measurements under shear, the home-made plane-plane shear cell described in \cite{tamborini_plasticity_2014} was used. The cell consists of two glass plates confining the sample in a gap ranging from $e=300 \um$ to $e = 1500 \um$. To reduce slip, the inner surfaces of the glass plates are frosted, leaving a small transparent window (of surface a few $\mathrm{mm}^2$) in order to optically probe the sample during deformation.
The deformation was imposed and measured as for the 2D samples. The typical accuracy on the strain measurement is of the order of $0.01\%$.

\subsection{Small-angle DLS apparatus}

\begin{figure}[h]
\centering
  \includegraphics[width=\columnwidth,clip]{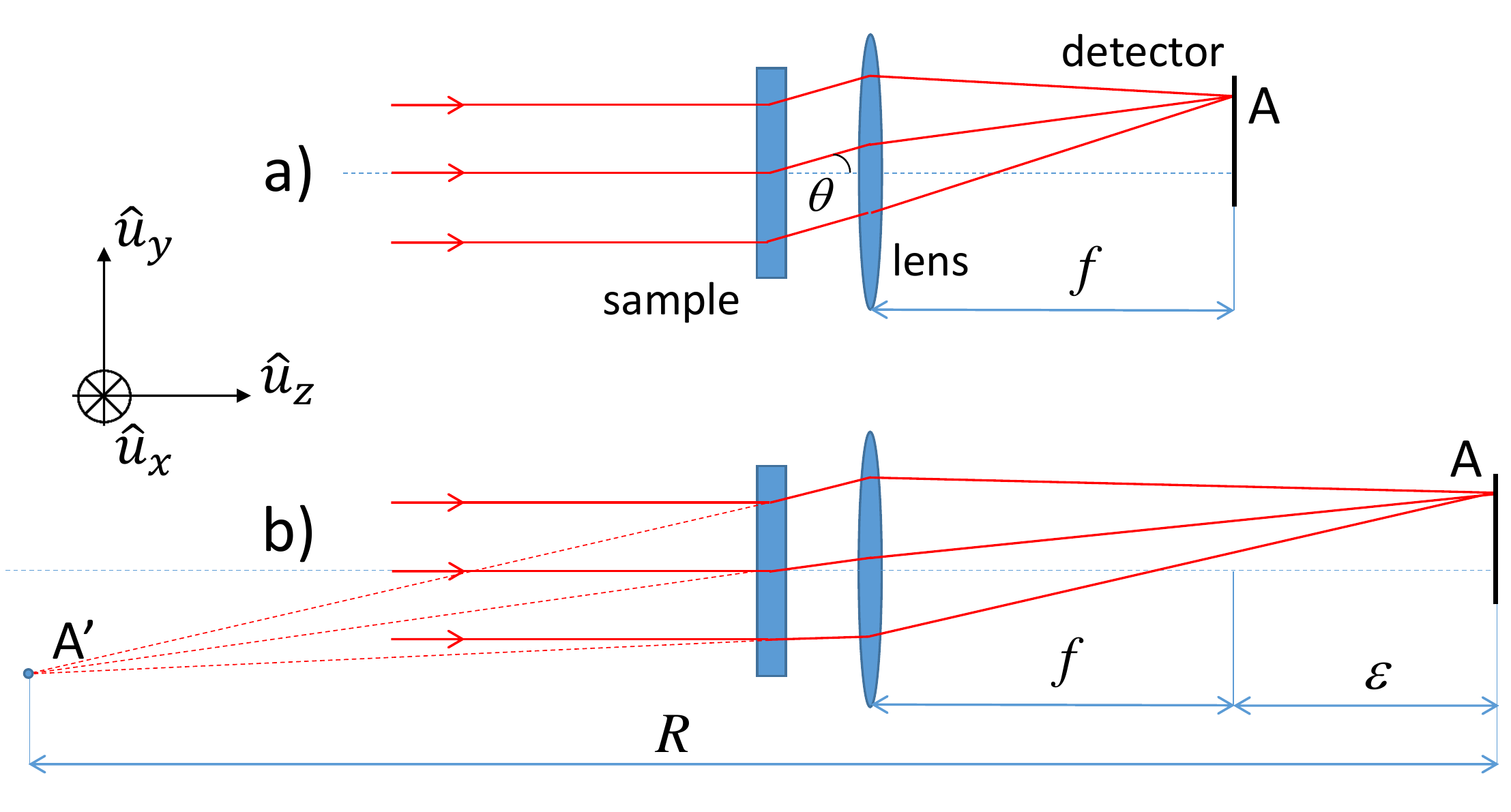}
  \caption{a): simplified scheme of the light scattering setup. The detector is placed in the focal plane of the Fourier lens. b): scheme used in the discussion of the effect of aberrations. $A$ and $A'$ are conjugated points, separated by a distance $R$ along the direction of the optical axis. The distance $\varepsilon$ has been exaggerated for the sake of clarity, usually $\varepsilon \ll f$.}
  \label{fig:setup}
\end{figure}

Dynamic light scattering experiments are performed using the custom-made apparatus described in~\cite{tamborini_multiangle_2012}, allowing simultaneous measurements at scattering angles in the range $0.4~\mathrm{deg} \le \theta \le 25~\mathrm{deg}$, corresponding to scattering vectors $q = 2 k \sin \theta/2$ in the range $0.1\um^{-1} \le q \le 5 \um^{-1}$, where $k = 2\pi n \lambda^{-1}$ is the wave vector of the incoming beam, with $n$ the refractive index of the solvent and $\lambda = 0.633~\um$ %\textbf{XXX Stefano check XXX}
the in-vacuo wave length of the laser source. A simplified scheme of the DLS apparatus is shown in Fig.~\ref{fig:setup}a). The setup uses the typical far-field configuration, where the detector (the 2D sensor of a CMOS camera) is placed in the focal plane of a lens collecting the light scattered by the sample, often termed the Fourier lens. In this configuration, light scattered at the same $\theta$ and the same azimuthal angle (with respect to the direction $\hat{u}_z$ of the incoming beam) is conveyed to a single point on the detector, irrespective of the location of the scatterers in the sample. %In particular, it is interesting to mention that scattering vectors with the same modulus $q$ but different azimuthal orientations can be decoupled by selecting two different ROIs equally distant from the $\bm{q}=0$ position. In this way, for a given $q$ value, we can selectively detect the signal scattered in the direction parallel to the shear direction (we will call this scattering vector $q_x$) or in the perpendicular ($q_y$) direction.
The incident beam has a Gaussian shape, with a $1/e^2$ radius $w=0.45~\mathrm{mm}$ at the sample position, and radius $w_0 = 0.444~\mathrm{mm}$ in the beam waist, where the radius is minimal~\cite{born_principles_2013}. The focal length of the Fourier lens is $f = 17.3~\mathrm{mm}$.

%------------------------------------------------------------------- FIG
%------------------------------------------------------------------ FIG

The images collected by the detector have a distinctive grainy appearance, with dark and bright spots termed speckles, arising from the interference of the photons scattered by the sample~\cite{goodman_speckle_2007}. In order to extract information on the dynamics, we quantify the temporal fluctuations of the speckles by calculating the degree of correlation between a pair of images taken at time $t$ and $t+\tau$~\cite{berne_dynamic_1976,duri_time-resolved-correlation_2005}:
%+++++++++++++++++++++++++++++++++++++++++++++++
\begin{equation}
    g_2(\bm{q}, t, \tau) - 1 = \frac{\left< I_p(t) I_p(t + \tau) \right>_{p\in ROI(\bm{q})}}{\left< I_p(t) \right>_{p\in ROI(\bm{q})} \left< I_p(t + \tau) \right>_{p\in ROI(\bm{q})}}-1 \,.
    \label{eqn:CorrelationDegree}
\end{equation}
%+++++++++++++++++++++++++++++++++++++++++++++++
Here, $I_p(t)$ is the intensity at time $t$ of the $p-$th pixel and the average is over a set of pixels or region of interest (ROI) corresponding to a well-defined $\mathbf{q}$ vector. For the optical layout sketched in Fig.~\ref{fig:setup} and $\mathbf{q} = q_x \hat{u}_x$ (respectively, $\mathbf{q} = q_y \hat{u}_y$), the ROI is centered around the position ($f\tan \theta~\hat{u}_x, 0$) (respectively, around the position ($0,f\tan \theta~\hat{u}_y $), see Ref.~\cite{tamborini_multiangle_2012} for more details.

Throughout this paper, the loss of correlation will be due exclusively to the imposed shear deformation or sample translation: the temporal variable $\tau$ in Eq.~\ref{eqn:CorrelationDegree} will be then replaced by $\gamma$ or $\delta$, respectively.
The intensity correlation function is related to the field correlation function by the Siegert relation~\cite{berne_dynamic_1976}, which for the case, e.g., of a sheared 3D sample reads:
%+++++++++++++++++++++++++++++++++++++++++++++++
\begin{equation}
    g_2(\bm{q}, \gamma)-1 = |g_1(\bm{q}, \gamma)|^2 \propto \left|\left<\bm{E}_S(\bm{q}, 0)\cdot\bm{E}_S^*(\bm{q}, \gamma)\right>\right|^2\,,
    \label{eqn:FieldCorrelation_def}
\end{equation}
%+++++++++++++++++++++++++++++++++++++++++++++++
where the (complex) scattered electric field is given by
%+++++++++++++++++++++++++++++++++++++++++++++++
\begin{equation}
    \mathbf{E}_S(\bm{q}, \gamma) \propto \sum\limits_{j=1}^N \bm{E}_{in}(\bm{r}_j(\gamma))e^{-i\bm{q}\cdot\bm{r}_j(\gamma)} \,,
    \label{eqn:scattered}
\end{equation}
%+++++++++++++++++++++++++++++++++++++++++++++++
with $\bm{r}_j$ the ($\gamma$- or $\delta$-dependent) coordinates of the $j-$th particle and where the sum runs over $N$ scatterers. In writing Eq.~\ref{eqn:scattered}, we have assumed with no loss of generality that all particles are identical and we have neglected any $q$ dependence of the scattering from an individual particle, i.e. we have set to unity the form factor~\cite{berne_dynamic_1976}. Note that in contrast to the usual expression of $\mathbf{E}_S$ we have included the possibility that the incoming electric field $\bm{E}_{in}$ varies spatially (both in phase and amplitude), to account for deviations with respect to illumination by a perfect, infinitely extended plane wave.

\subsection{Numerical simulations}

Numerical simulations of the scattering signal associated to a given sample deformation were performed by generating a 3D random set of scatterers with the same size as the experimental sample and by propagating the scattered field from the sample to the detector plane, where $\mathbf{E}_S$ is obtained from Eq.~\ref{eqn:scattered}. For each pixel location, the scattered intensity is then calculated as $|\mathbf{E}_S|^2$; the simulated speckle patterns are then analyzed as the experimental ones, i.e. using Eq~\ref{eqn:CorrelationDegree}. For the experimental data to be easily compared to theory and simulations, all correlation functions were normalized such that $g_2(\bm{q}, 0)-1 = 1$.

\subsection{Measurements of non-affine displacements by optical microscopy}

For checking purposes, non-affine displacements of the $\mathrm{TiO}_2$ tracer particles in the PA gels were also measured in real space by coupling the shear cell to a bright field microscope, whose condenser diaphragm was fully open in order to reduce the depth of focus. The in-plane non-affine displacements were measured for different depths $z$ and averaged over all particles. We use a custom python code built from the Trackpy Python package ~\cite{allan_trackpy:_2015} to track the colloidal particles. The non-affine displacements measured by microscopy are compared to those obtained by DLS. Additionally, they are used to obtain a reference value for numerical simulations, where non-affinity is introduced by adding to the affine displacement field a random, isotropic and Gaussian-distributed extra-contribution, in such a way that the resulting rms displacement matches the one measured by microscopy.

\section{Dynamic Light Scattering for a sheared sample}
\label{sec:DLSresults}

Equation~\ref{eqn:scattered} shows that all scatterers contribute to the signal detected in a DLS experiment via phase terms: this is what makes DLS such a powerful and sensitive technique to probe the sample dynamics. Indeed, every change in the ${\bm{r}_j}$ coordinates has an impact on the scattered field, and thus on the correlation function, Eq.~\ref{eqn:CorrelationDegree}. Under a shear deformation, the affine displacement field produces a loss of correlation that can be calculated from Eq.~\ref{eqn:FieldCorrelation_def} and the Siegert relation. In the presence of non-affine displacements, the particles' displacement contains and additional term: $\Delta \mathbf{r}_j = \Delta \mathbf{r}^{(\mathrm{aff})}_j + \bm{R^\prime}_j$. The non-affine contribution $\bm{R^\prime}$ results in a different, generally faster decay of $g_2-1$, as compared to the case of a purely affine deformation. In this section we will show how to decouple the two contributions, by quantifying the average non-affine displacement and `filtering out' the decorrelation due to the affine deformation.

To this end, it is useful to proceed by steps. Equation~\ref{eqn:AffineDeformation} indicates that a 3D sample deformed affinely may be modeled by  a set of $\Sigma_z$ planes perpendicular to the optical axis, rigidly translating by a $z$-dependent amount $\bm\delta_z=\gamma z \hat{u}_x$. The decay of $g_2-1$ will then contain a first contribution due to the rigid translation of each plane, plus a second contribution arising from the relative motion of different $\Sigma_z$ slices. We will thus start by considering the simple case of a 2D sample that rigidly translates along the $x$ axis, discussing all the factors that contribute to the loss of correlation (Sec.\ref{subsec:DLStrasl}). We will then address the more complicated situation of a 3D sample composed of a stack of $\Sigma_z$ slices (Sec. \ref{sec:DLS_affine}). Finally, in Sec.~\ref{sec:DLS_nonaff} we will discuss the general case of both affine and non-affine displacements.

\subsection{Rigid translation of a 2D sample}
\label{subsec:DLStrasl}

We model a 2D sample by a set of scatterers with positions $\bm{r}_j$ such that $\bm{r}_j \cdot \hat{u}_z = z = const$ and refer to it as to a $\Sigma_z$ plane. We consider the case of a rigid translation $\bm\delta=\gamma z \hat{u}_x$ of such plane. If the incident beam was an ideal plane wave, infinitely extended in the $\hat{u}_x, \hat{u}_y$ directions perpendicular to the propagation direction, the term $\bm{E}_{in}$ in Eq.~\ref{eqn:scattered} would be space-independent and could be factored out of the sum. It is then easy to show that the intensity correlation function $g_2-1$ would be invariant under a sample translation, since the total electric field would change only by a phase factor $\exp\left(-i\bm{q}\cdot\bm\delta\right)$ as the sample drifts. This is why DLS is commonly regarded as a technique sensitive to the relative motion of the scatterers, not to their global drift. However, ideal plane wave conditions cannot be fully met in experiments: the incoming beam has a finite cross section, often with a radial intensity modulation, e.g. a Gaussian profile. In this section we thus discuss the decorrelation expected for rigid translations under a non-uniform incident electric field.

Intuitively, it is clear that translations larger than the beam size lead to the full decay of $g_2$, since they completely change the set of illuminated scatterers. Even when the sample is translated by an amount smaller than the scattering volume diameter, some loss of correlation is expected if the sample illumination is not uniform: as the sample is translated, each scatterer receives a varying illuminating field, which will modify the relative weight of the terms in the sum of Eq.~\ref{eqn:scattered} and thus the scattered intensity. To quantify this effect, we consider the realistic case of a Gaussian laser beam propagating along the $z$ axis in the paraxial approximation \cite{SveltoLasers}:
%+++++++++++++++++++++++++++++++++++++++++++++++
\begin{equation}
    \bm{E}_{in}(r, z) = \bm{E}_0 \frac{w_0}{w(z)}e^{ikz}e^{-\frac{r^2}{w^2(z)}}e^{ik\frac{r^2}{2\rho(z)}}
    \label{eqn:GaussianBeam}
\end{equation}
%+++++++++++++++++++++++++++++++++++++++++++++++
where $(r, z)$ are radial and axial cylindrical coordinates along the optical axis, $w(z)=w_0\sqrt{1+(z/a)^2}$ is the beam $1/e^2$ radius and $\rho(z)=z[1+(a/z)^2]$ is the radius of curvature of the wavefront, as a function of the position $z$ along the optical axis. Here $a=\frac{1}{2}kw_0^2$ is the depth of the Rayleigh region, i.e. the region centered around $z=0$ where the laser beam can be considered approximatively plane and collimated.

Using this expression in Eqs.~\ref{eqn:scattered} and \ref{eqn:FieldCorrelation_def}, the intensity correlation function is found to be:
%+++++++++++++++++++++++++++++++++++++++++++++++
\begin{equation}
     g_2(\bm{q}, \bm\delta) - 1 = e^{- \frac{\left| \bm\delta \right|^2}{w^2}\left[1 + \left(\frac{kw^2}{2\rho}\right)^2\right]}
    = e^{- \frac{\left| \bm\delta \right|^2}{w_0^2}}
    \label{eqn:IntensityCorrelation}
\end{equation}
%+++++++++++++++++++++++++++++++++++++++++++++++
with $\bm\delta$ the translation vector, along the $x$ axis. As anticipated, the intensity pattern decorrelates significantly when the 2D object translation is comparable to the beam size. Interestingly, Eq.~ \ref{eqn:IntensityCorrelation} shows that the relevant length scale is $w_0$, the beam size at the beam waist ($z=0$), regardless of the actual $z$ position of the sample. This is because the increase of the beam size $w(z)$ and the decrease of the radius of curvature $\rho(z)$ as the sample departs from the beam waist exactly compensate each other, leading to the simple expression in the far r.h.s. of Eq.~\ref{eqn:IntensityCorrelation}.
%------------------------------------------------------------------- FIG
\begin{figure}[h]
\centering
  \includegraphics[width=\columnwidth,clip]{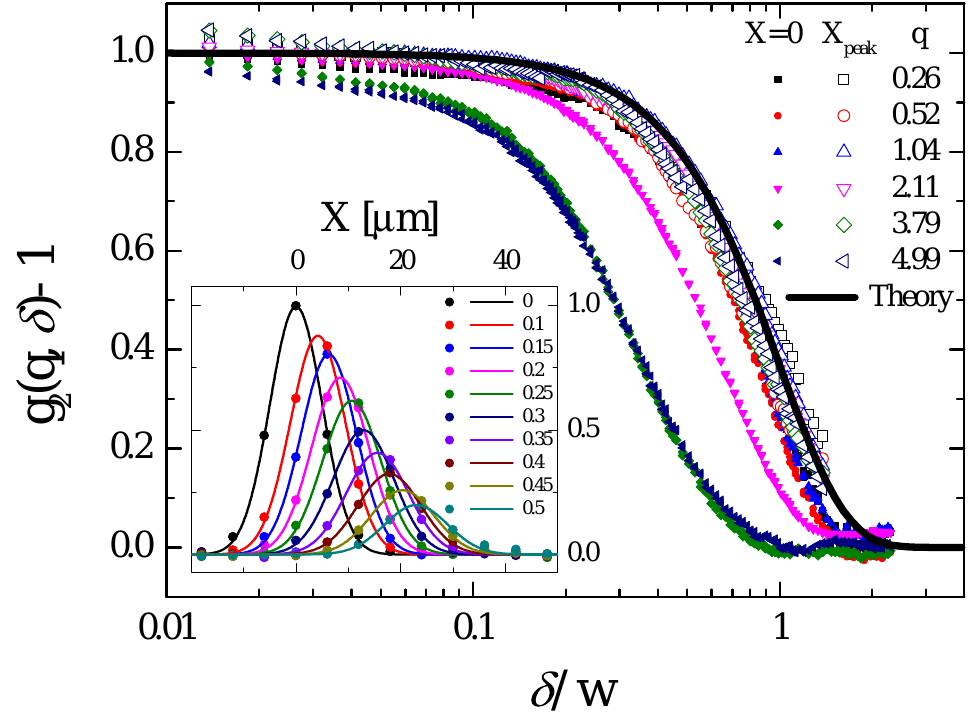}
  \caption{Main plot: decay of the intensity correlation function due to the rigid translation of a quasi-2D object, measured at different scattering vectors $q$, as indicated by the label (units: $\um^{-1}$). Filled symbols: usual $g_2-1$, calculated pixel-by-pixel following Eq.~\ref{eqn:CorrelationDegree}. Open symbols: correlation function as obtained from the height of the peak in the spatio-temporal crosscorrelation, see text for details. Black line: Eq.~\ref{eqn:CrossCorr_ref} evaluated for $\bm{X}=\bm{X}_{peak}=\sigma\bm{\delta}$.
Inset: cut of the intensity spatio-temporal crosscorrelation, Eq.~\ref{eqn:CrossCorr_ref}, along the translation direction $\hat{u}_x$, plotted for different values of $\delta$, as indicated by the legend (in mm), and for $q=3.79 \um^{-1}$. Symbols: experimental data; lines: fits via Eq.~\ref{eqn:CrossCorr_ref}.}
  \label{fig:2D_trasl}
\end{figure}
%------------------------------------------------------------------ FIG

According to Eq.~\ref{eqn:IntensityCorrelation}, the decay of $g_2-1$ should not depend on the scattering vector. However, Fig.~\ref{fig:2D_trasl} shows that the experimental correlation functions decay increasingly rapidly at large $q$. We now show that this $q$-dependence is due to the optical aberrations and deviations from the paraxial approximation, which become increasingly important at large $q$. As far as speckle decorrelation is concerned, the most relevant effect is the fact that the focal surface is curved and not planar, as in the paraxial, aberrations-free approximation. Thus, it is impossible that all points of a flat detector lay in the focal surface. The situation is schematized in Fig.~\ref{fig:setup}b), where $\varepsilon$ is the offset along the $\hat{u}_z$ direction of the detector point A, with respect to the focal surface. Figure~\ref{fig:setup}b) shows that under these conditions the point A collects light issued by the whole illuminated sample, but with a scattering angle that slightly depends on the scatterer position. The scattered light collected in A appears to originate from a point A' located before the sample and conjugated to A by the Fourier lens. The $z$ component of the distance AA', denoted by $R$, plays a key role in the following.  Asakura and Takai~\cite{asakura1981} have studied mixed spatio-temporal intensity correlation functions describing the evolution of the speckle pattern generated by a translating object placed at an arbitrary distance from the detector. Their key result was that a sample translation results in general in a combination of a translation of the speckle pattern and random fluctuations of the speckle intensity (`boiling'). We thus expect both contributes to exist in our experiment, unlike the ideal case of a detector placed exactly in the focal surface and a sample in the beam waist, for which the speckle pattern is reconfigured with no overall drift.

We adapt the results of Ref.~\cite{asakura1981} to the geometry described here, and define a spatio-temporal intensity crosscorrelation $g_2\left(\bm{X}, \bm{\delta}\right) - 1 \propto \left|\left<\bm{E}_S\left(\bm{r}, 0\right)\cdot\bm{E}_S^*\left(\bm{r}+\bm{X}, \bm\delta\right)\right>\right|^2$, where $\bm{r}$ and $\bm{X}$ are vectors in the sensor plane. For a translating sample illuminated by a Gaussian beam (Eq.~\ref{eqn:GaussianBeam}), one finds:
%+++++++++++++++++++++++++++++++++++++++++++++++
\begin{equation}
     g_2(\bm{X}, \bm\delta) - 1
    = e^{-\frac{\left| \bm\delta \right|^2}{w^2}}e^{-\frac{\left| \bm{X} - \sigma \bm\delta \right|^2}{\Delta^2}}
%    = e^{-\frac{\left| \mathbf\delta - \frac{\alpha}{\sigma} \mathbf{X} \right|^2}{l^2}}e^{-\frac{\left| \mathbf{X} \right|^2}{\Delta^2} (1-\alpha)}
    \label{eqn:CrossCorr_ref}
\end{equation}
%+++++++++++++++++++++++++++++++++++++++++++++++
where $\Delta=R/(kw)$ is the speckle size, $\sigma=1+R/\rho$, and $\bm{X}$ the relative position of two points on the sensor whose intensity is crosscorrelated. For $\bm{X}=0$, Eq.~\ref{eqn:CrossCorr_ref} reduces to the standard pixel-by-pixel intensity correlation function, Eq.~\ref{eqn:FieldCorrelation_def}. In this case, one retrieves $g_2-1=\exp(-\delta^2/l^2)$, similarly to Eq.~\ref{eqn:IntensityCorrelation}, but with a faster decay, since $w_0$ is replaced by a shorter characteristic length $l$ given by
%+++++++++++++++++++++++++++++++++++++++++++++++
\begin{equation}
    \frac{1}{l^2} = \frac{1}{w^2} + \frac{\sigma^2}{\Delta^2} = \frac{1}{w_0^2}\left\{ 1 + \left(\frac{a}{R}\right)^2\left[1+\left(\frac{z}{a}\right)^2\right]\left(1+\frac{2R}{\rho}\right)\right\}
    \label{eqn:ModifiedCorrelationLength_Article}
\end{equation}
%+++++++++++++++++++++++++++++++++++++++++++++++

The ideal case of far field scattering (achieved when the detector images the focal surface of the Fourier lens) corresponds to the $R\rightarrow\infty$ limit, where we retrieve as a characteristic length $l=w_0$, as in Eq.~\ref{eqn:IntensityCorrelation}. On the other hand, a non-vanishing deviation $\varepsilon$ of the detector position with respect to the focal surface can be described in terms of a finite value $R(\varepsilon) \approx f^2/\varepsilon$, which results in a faster decorrelation. Both optical aberrations (embedded in $\sigma$, via $R$) and wavefront curvature (embedded again in $\sigma$, via $\rho$) tend to enhance the system sensitivity to rigid translations, which is not desired if one aims to detect non-affine rearrangements. Moreover, equation \ref{eqn:CrossCorr_ref} suggests that both effects have the same physical nature: as a consequence of either one, a collective drift motion superimposes to the normal, `boiling' decorrelation of the speckle pattern. Indeed, when $\rho$ or $R$ assume finite values, it is easy to see from Eq.~\ref{eqn:CrossCorr_ref} that the maximum correlation is achieved for a non-null $\bm{X}$ value, which reflects the presence of a global drift that adds up to the `boiling' decorrelation (see inset of Fig.~\ref{fig:2D_trasl}). To correct the data for this effect, one can follow the collective speckle translation by tracking the position $\bm{X}_{peak}=\sigma\bm{\delta}$ of the crosscorrelation peak. The drift-corrected value of $g_2-1$ is then taken as the height of the peak, i.e. the value of the spatio-temporal crosscorrelation for $\bm{X} = \bm{X}_{peak}$. Following this strategy, one finds that the corrected $g_2-1$ decays with the sample drift as
\begin{equation}
g_2(\bm{X}_{peak}, \delta)-1=\exp(-\delta^2/w^2) \,.
\end{equation}
The open symbols in the main graph of Fig.~\ref{fig:2D_trasl} show the drift-corrected intensity correlation function: in contrast to the usual pixel-by-pixel correlation function (solid symbols), the decay of $g_2-1$ with sample displacement $\delta$ is $q$-independent and very well accounted for by Eq.~\ref{eqn:CrossCorr_ref} (line).

To summarize, in this section we have shown that the finite size of the beam, the curvature of the wavefront impinging on the sample, and optical aberrations all contribute to the decay of the intensity correlation function upon translating a 2D sample. When correcting for the speckle drift, there is no $q$ dependence in these effects. Finally, the decay of the correlation function is negligible for $\delta << w$, i.e. when the sample translation is (much) smaller than the beam size.

\subsection{Affine deformation of a 3D sample}
\label{sec:DLS_affine}

The case of a 3D object undergoing an affine deformation can be treated by decomposing the sample in slices, each of which translates by a $z$-dependent amount. The decay of $g2-1$ contains a first contribution due to the rigid translation of each slice, as well as a second contribution arising from the relative motion of scatterers belonging to distinct slices.
In the ideal case where an infinite plane wave illuminates the sample and the detector is in the focal plane, the first contribution does not lead to a loss of correlation, while the second one is readily computed (see Appendix 2 for details). One finds
%+++++++++++++++++++++++++++++++++++++++++++++++
\begin{equation}
    g_2(\bm{q}, \gamma) - 1 = \sinc^2\left( \gamma q_x \frac{e}{2} \right)
    \label{eqn:IntensityCorrelation_Largeq}
\end{equation}
%+++++++++++++++++++++++++++++++++++++++++++++++
where $q_x=\bm{q}\cdot\hat{u}_x$ is the component of the scattering vector parallel to the shear direction and $\sinc(x) = x^{-1} \sin(x)$. Equation~\ref{eqn:IntensityCorrelation_Largeq} illustrates the ability of DLS to selectively measure affine or non-affine displacements, as mentioned in Sec.~\ref{sec:introduction}. Indeed, if the azimuthal orientation of the scattering vector is  chosen such that $q_x = 0$, $g_2-1$ is insensitive to affine displacements. Any decay of the correlation function that may be observed is then to be ascribed to non-affine dynamics. The scheme of Fig.~\ref{fig:setup}a) illustrates a geometry where the scattering vector is orthogonal to the shear direction: the scattered rays belong to the $(y,z)$ plane and hence $q_x=0$ (we remind that $\mathbf{q} = \mathbf{k}_{sc}-\mathbf{k}_{in}$, with $\mathbf{k}_{in}$ and $\mathbf{k}_{sc}$ the wave vector of the incident and scattered light, respectively). Note that the loss of correlation depends on the variable $\gamma e$, i.e. the relative displacement of the two plates confining the sample. Thus, for a given $\gamma$, thinner samples exhibit a smaller loss of correlation.
%------------------------------------------------------------------- FIG
\begin{figure}[h]
\centering
  \includegraphics[width=1\columnwidth,clip]{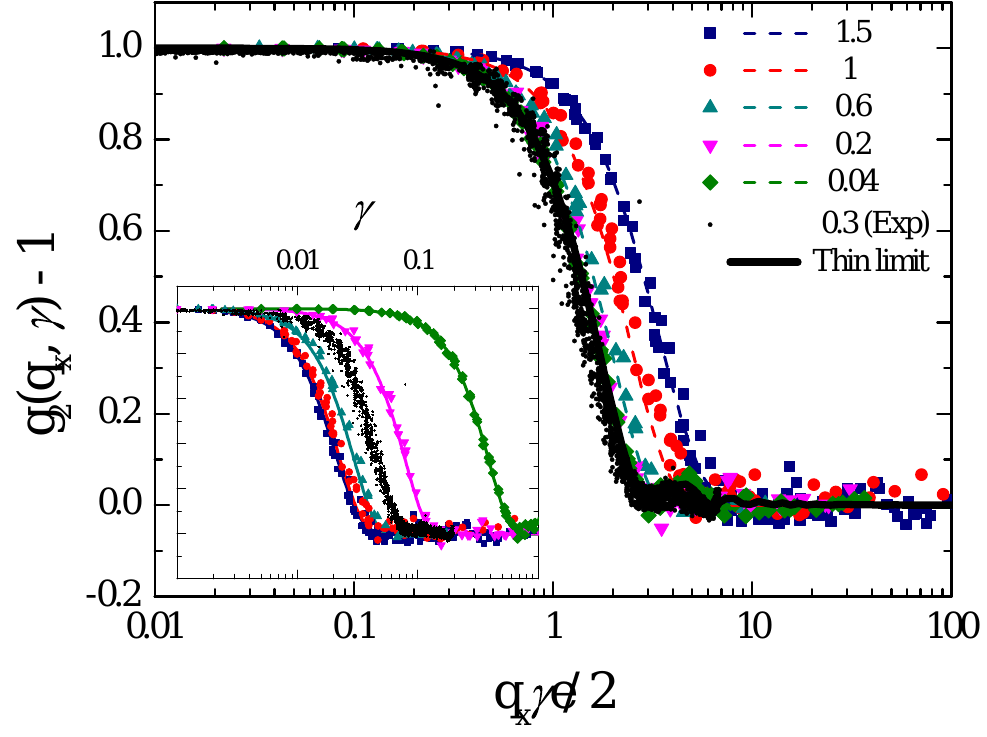}
  \caption{Inset: $g_2-1$ for a 3D sample, as a function of the imposed shear deformation, in the $q_x \gg \frac{e}{\gamma w^2}$ limit, for $q_x = 1.04 \um^{-1}$. Large colored symbols: numerical simulations for sample thickness from 0.04  mm to 1.5 mm, as shown by the label of the main plot. Lines: numerical integration of Eq.~\ref{eqn:FieldCorrelationShear}. Black dots: experimental data for $e=0.3$ mm. Main plot: same data plotted as a function of the scaled variable $q_x \gamma \frac{e}{2}$. Black line: theoretical curve in the $e \ll w$ limit, Eq.~\ref{eqn:IntensityCorrelation_Largeq}. The beam size used in simulations is $w=450 \um$, the same as in experiments. For $e < w$ both experimental data and simulations follow Eq.~\ref{eqn:IntensityCorrelation_Largeq}, whereas deviations are observed for thicker samples, in good agreement with theoretical expectations.}
  \label{fig:3D_shear_qparall}
\end{figure}
%------------------------------------------------------------------ FIG

When the finite beam size and wavefront curvature are taken into account, the expression for $g_2-1$ or, equivalently, for $g_1 = \sqrt{g_2-1}$, become more complicated, but remains independent of $q_y$. In Appendix 2 we provide an expression for $g_1$ for a sample in the waist of a Gaussian beam (see Eq.~\ref{eqn:FieldCorrelationShear} in Appendix 2). The thin sample ($e \ll w$) and high-$q_x$ ($q_x \gg \frac{e}{\gamma w^2}$) limits of this expression reduce to the simpler Eq.~\ref{eqn:IntensityCorrelation_Largeq}. We test these expressions for a scattering vector parallel to the shear direction in Fig.~\ref{fig:3D_shear_qparall}. The inset shows $g_2-1$ as a function of the imposed shear deformation $\gamma$, for simulations (large symbols) and for one experimental run on a PA gel (small black dots). The lines are the result of the numerical integration of Eq.~\ref{eqn:FieldCorrelationShear} in Appendix 2: an excellent agreement is observed between theory and simulations. The main plot shows the same data, plotted vs the scaled variable $q_x \gamma e /2$. The experimental data and the simulations for $e \le 200 \um$ fall onto a single master curve, well reproduced by the simple form of Eq.~\ref{eqn:IntensityCorrelation_Largeq} (line). Additional experimental data for several scattering vectors in the range $0.1\um^{-1}<q_x<4\um^{-1}$ also fall onto the same master curve (data not shown to avoid overcrowding the plot). Since here $w = 450 \um$, this indicates that the thin sample expression mathematically derived for $e \ll w$ actually remains valid up to the somehow more relaxed regime $e \lesssim w$. For $e > w$, by contrast, $g_2-1$ significantly departs from Eq.~\ref{eqn:IntensityCorrelation_Largeq} and the numerical integration of the full expression, Eq.~\ref{eqn:FieldCorrelationShear}, is required to account for the simulations.

%------------------------------------------------------------------- FIG
\begin{figure}[h]
\centering
  \includegraphics[width=1\columnwidth,clip]{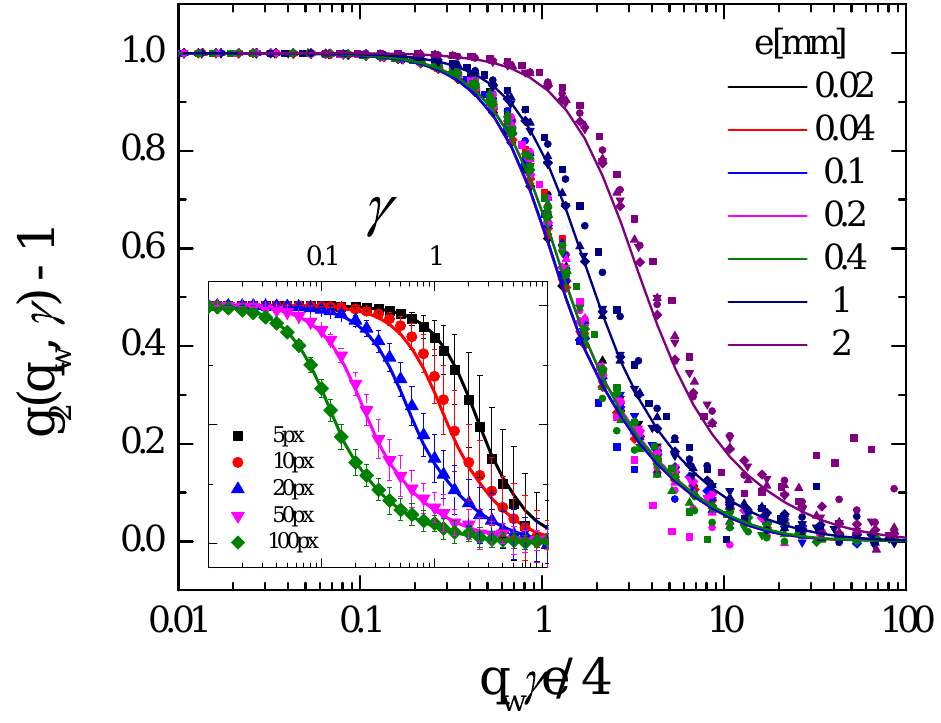}
  \caption{Inset: effect of the finite $q_x$ component of the scattering vectors associated to the ROI chosen for calculating $g_2-1$. Symbols: simulations for a purely affine deformation of a 3D sample. Lines: numerical integration of Eq.~\ref{eqn:FieldCorrelationShear} over the relevant $q_x$ range, indicated in the legend, in units of pixels. In all cases, the gap is $e = 200 \um$. Main plot: Effect of the finite size of the ROI, for several sample thicknesses, as specified in the legend. Different symbols refer to various ROI sizes: 5 px (squares), 10 px (circles), 20 px (triangles), 50 px (down triangles), 100 px (diamonds). See the text for the definition of $q_w$, the ROI half width in Fourier space.}
  \label{fig:ROI_thickness}
\end{figure}
%------------------------------------------------------------------ FIG

A key point that has to be taken into account in realistic simulations and in experiments is the finite size of the ROIs over which the intensity correlation function is averaged (see Eq.~\ref{eqn:CorrelationDegree}). Since the ROIs must contain a sizeable number of speckles in order to achieve a good statistics, it is impossible to measure $g_2-1$ for a $q$ vector exclusively oriented in the $\hat{u}_y$ direction: any viable ROI will correspond to a range of $q$ vectors with a finite $x$ component. This potentially limits the feasibility of the strategy outlined above for measuring non-affine displacements, which in principle requires to acquire $g_2-1$ for a $q$ vector with no component along the shear direction $\hat{u}_x$.

To explore this issue, we show in Fig.~\ref{fig:ROI_thickness} correlation functions obtained from simulations, where the data have been analyzed using a ROI corresponding to $q_y$ in the range $0.1 - 10 \um^{-1}$ and various $q_x$, expressed here in terms of the ROI size along $\hat{u}_x$, in pixels (1 pixel corresponds indicatively to $5 \times 10^{-3} \um^{-1}$). The results are independent of $q_y$: we thus show data averaged over all $q_y$. The curve for the thinnest ROI (5 pixel along $\hat{u}_x$) is almost flat up to $\gamma \approx 1$. This demonstrates that under realistic conditions one can indeed measure a correlation function that is sensitive essentially only to non-affine displacements, up to a shear deformation of order one. This comes however at the expenses of statistics: since the ROI is relatively small, large fluctuations are seen in the data, as shown by the error bars, which quantify the standard deviation of $g_2-1$ over 20 independent runs. As the size of the ROI along $\hat{u}_x$ is increased, the statistics improves significantly, but $g_2-1$ start decaying at smaller strains. The lines are theoretical expectations obtained by numerically averaging Eq.~\ref{eqn:FieldCorrelationShear} over the range of $q_x$ associated to the ROI. They are in excellent agreement with the simulations, indicating that the simple theory developed here can be used to quantitatively predict the impact of the ROI width, thereby providing valuable guidance for the optimization of the analysis parameters.

Based on Eq.~\ref{eqn:FieldCorrelationShear}, one expects that the impact of the ROI width on $g_2-1$ varies also with the sample thickness $e$. The main plot of Fig.~\ref{fig:ROI_thickness} rationalizes the dependence on $e$, showing that data for different widths and a given cell gap fall onto the same master curve, provided that $g_2-1$ is plotted against a scaled shear deformation, $q_w \gamma e /4$. Here, $q_w$ quantifies the width of the ROIs, which extend from $-q_w$ to $q_w$ in the $\hat{u}_x$ direction. The shape of these master curves depend on $e$; remarkably, they collapse on top of each other in the thin sample limit. Again, a very good agreement is found between the correlation functions obtained by simulations and their theoretical expression (lines in the main plot of Fig.~\ref{fig:ROI_thickness}).

To recapitulate the main findings for DLS under a purely affine displacement field, we have shown that the intensity correlation function decays faster (i.e. at smaller strains) for thicker samples. In the limit $e \lesssim w$ the effects of the finite beam size are negligible. Under these conditions, correlation functions for $\mathbf{q}$ parallel to the shear direction depend only on the scaled variable $e \gamma$. In experiments and realistic simulations, the finite $x$ component of the scattering vectors associated to a ROI has to be taken into account, even when the ROI corresponds essentially to the direction orthogonal to the applied shear. This (albeit small) $x$ component is responsible for the decay of $g_2-1$, which can be rationalized using the scaled variable $q_w e \gamma$. Finally, we emphasize that for the purely affine deformation discussed so far, the dynamics are always independent of $q_y$.

\subsection{Probing non-affine displacements}
\label{sec:DLS_nonaff}

\begin{figure}[h]
\centering
  \includegraphics[width=\columnwidth,clip]{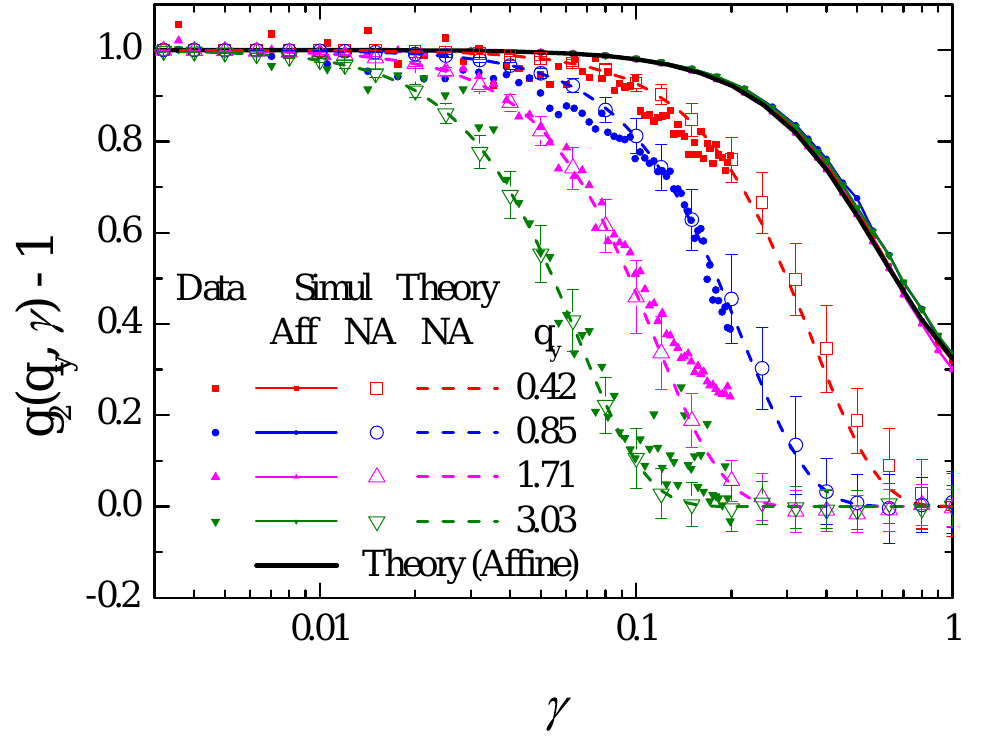}
  \caption{Strain dependence of the intensity correlation function for a sheared PA gel. Small filled symbols: experimental data for several $q_y$ values, as indicated by the labels (units: $\um^{-1}$). Thick black line: theoretical prediction for a purely affine deformation, obtained by the numerical integration of Eq.~\ref{eqn:FieldCorrelationShear}. Small solid symbols connected by lines: $g_2-1$ obtained from simulations, for a purely affine deformation. Theory and simulations deviate strongly from the experimental data, indicating that non-affine displacements must be included in the modelling. Dashed lines and large open symbols: correlation functions obtained from theory (Eq.~\ref{eqn:NonaffineParameter}) and simulations, respectively, assuming Gaussian-distributed non-affine particle displacements, with a rms value given by Eq.~\ref{eqn:NonaffineParameter}. In simulations, data are averaged over $N=50$ independent runs and error bars represent the run-to-run standard deviation. In both theory and simulations, a generalized diffusion coefficient $c = 30 \um^2$ was used.}
  \label{fig:Affine_vs_nonaffine}
\end{figure}

In Fig.~\ref{fig:3D_shear_qparall} we have shown that experimental correlation functions for a PA gel, measured at $q$ vectors oriented parallel to the shear direction, agree well with numerical and theoretical predictions for a purely affine deformation. We now inspect data from the same experiment, but analyzed for a scattering vector oriented in the perpendicular direction $\hat{u}_y$. The results are shown in Fig.~\ref{fig:Affine_vs_nonaffine} (small solid points), together with the corresponding correlation functions obtained by theory and simulations using ROIs of the same size as in the experiment (width along $\hat{u}_x = 10$ pixels) and assuming a purely affine deformation (thick black line and small symbols connected by lines for theory and simulations, respectively). The experimental data deviate strongly from theory and simulations, both quantitatively and qualitatively: the decay of the experimental $g_2-1$ occurs at much smaller strains $\gamma$ and depends strongly on $q_y$.

%------------------------------------------------------------------- FIG
%------------------------------------------------------------------ FIG

\begin{figure}[h]
\centering
  \includegraphics[width=1\columnwidth,clip]{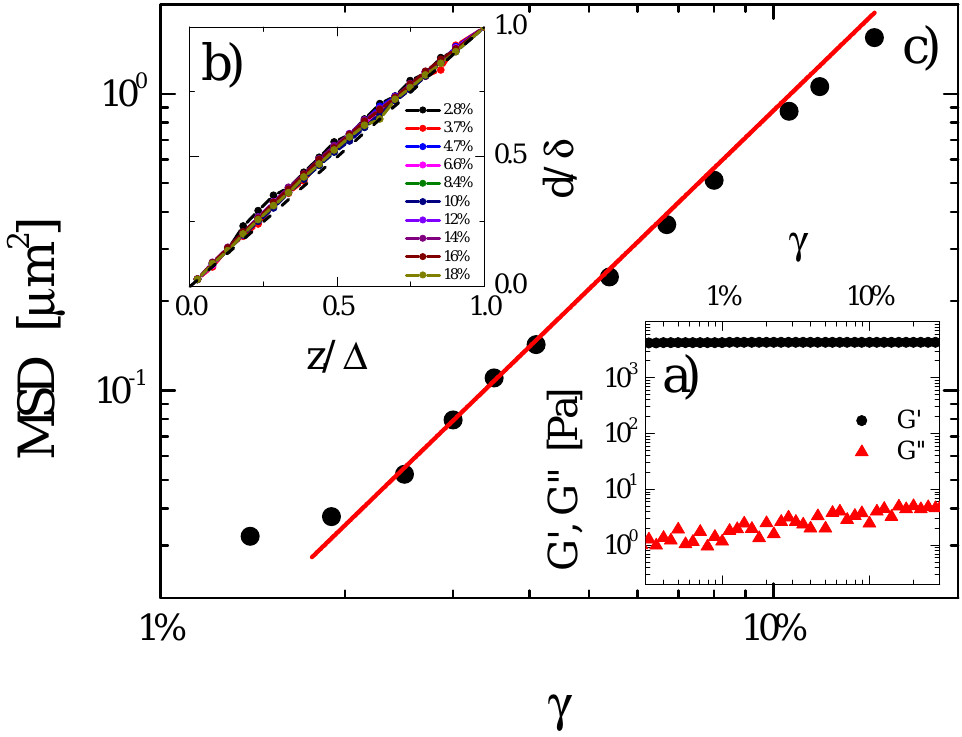}
  \caption{Microscopy and rheology experiments measuring non-affinity in a PA hydrogel. a): rheological response of the gel in a strain sweep oscillatory test, at a frequency of $1~\mathrm{Hz}$.  b): deformation profile $d(z)$, normalized by the displacement $\delta$ imposed to the upper plate of the shear cell. The distance $z$ to the stationary plate is normalized by the gap $e$. c): Non-affine parameter $\mathcal{A}$ as a function of the imposed strain, as obtained by tracking the displacement of tracer particles. A quadratic fit of $\mathcal{A}(\gamma)$ (line) yields a generalized diffusion coefficient $c=75 \pm15 \um^2 $. %\textbf{XXX STEFANO: provide error on $c$. Add a), b), c) in Inset 2, Inset 1 and main plot, respectively. Cut the shear profiles beyond 20\% deformation, where you observe slip. Indicate frequency used for the strain sweep. XXX}
}
  \label{fig:NonAffinity_tracking}
\end{figure}

We attribute this discrepancy to non-affine displacements in the PA gel that were not taken into account in the simulations nor in the theory. In Sec.~\ref{sec:introduction} we mentioned several mechanisms that may lead to a non-affine component of the displacement field. In the experiments, the applied strain is relatively modest, $\gamma \le 0.3$. Rheology measurements indicate that in this regime the sample response is essentially $\gamma$-independent (see Fig.~\ref{fig:NonAffinity_tracking}a)), suggesting negligible non-linear effects and plasticity. We thus propose that non-affinity stems from spatial fluctuations of the elastic shear modulus $G$, as reported in Ref.~\cite{basu_nonaffine_2011} for similar PA gels. Indeed, in real samples inhomogeneities are always present at a microscopic scale; from a mechanical point of view they can be described by local fluctuations of the shear modulus: $G(\bm{x}) = \overline{G} + \delta G(\bm{x})$, with $\overline{G}=\overline{G(\bm{x})}$ the spatially-averaged shear modulus and $\overline{\delta G(\bm{x})}=0$. Due to these fluctuations, the sample deformation locally deviates from affinity, originating a non-affine displacement field $\bm{R}^\prime(\bm{x})$ that adds up to the affine one. Note that we expect the non-affine component to be small in comparison to the affine one, since correlation functions measured in the shear direction are well reproduced by the affine component alone (see the small black dots in Fig.~\ref{fig:3D_shear_qparall}).

The non-affine displacement field may be quantified by its $\gamma$-dependent mean squared value, referred to as the non-affine parameter $\mathcal{A}$ in Ref.~\cite{basu_nonaffine_2011}, and related to the local shear modulus fluctuations. %, expressed in terms of its spatial correlation function: $\Delta^G(\bm{x}, \bm{x}^\prime) = \left<\delta G(\bm{x}) \delta G(\bm{x}^\prime)\right>$.
Under reasonable assumptions~\cite{didonna_nonaffine_2005,basu_nonaffine_2011}, %\textbf{XXX STEFANO: puoi ricordare le reasonable assumptions in due parole? XXX},
the non-affine parameter grows quadratically with the macroscopic deformation:
%+++++++++++++++++++++++++++++++++++++++++++++++
\begin{equation}
    \mathcal{A} = \frac{1}{N} \sum\limits_{j=1}^N \left|\bm{R'}_j\right|^2 %= \int\frac{d^3\bm{q}}{\left(2 \pi\right)^3} \frac{\gamma^2\Delta^G(\bm{q})}{q^2G^2}
    = c \gamma^2\,,
    \label{eqn:NonaffineParameter}
\end{equation}
%+++++++++++++++++++++++++++++++++++++++++++++++
%here $\Delta^G(\bm{q})$ is the Fourier transform of the spatial correlation function of $\delta G(\bm{x})$.
where the rms displacement per unit squared shear deformation, $c$, is a generalized diffusion coefficient, by analogy to Brownian diffusion originating from thermal fluctuations, with $\gamma^2$ playing the role of time in ordinary diffusion.

%------------------------------------------------------------------- FIG
%------------------------------------------------------------------ FIG

In order to test whether the decay of $g_2-1$ observed in the experiments for $q$ in the $\hat{u}_y$ direction can indeed by ascribed to non-affine displacements, we measure $\mathcal{A}$ in independent microscopy experiments, where the motion of the tracer particles is tracked while applying a shear deformation to the sample. Figure~\ref{fig:NonAffinity_tracking}b) shows that up to $\gamma = 0.18$ the deformation profile obtained by averaging the displacements of all particles at a distance $z$ from the immobile plate follows is linear, thus ruling out slip and shear banding. For larger strains, slip starts occurring close to the immobile plate (data not shown). Data with slip are disregarded in the calculation of $\mathcal{A}$; moreover, we assume that the non-affine displacements are isotropic and multiply the non-affine parameter obtained by 2D microscopy by a factor of 3/2, in order to obtain the 3D $\mathcal{A}$. Figure~\ref{fig:NonAffinity_tracking}c) shows $\mathcal{A}$ vs the imposed shear deformation in a double logarithmic plot. The quadratic law, Eq.~\ref{eqn:NonaffineParameter}, fits very well the data (line), yielding $c = 75 \pm 15 \um^2$.

To model the effect of non-affine displacements on the intensity correlation function, we assume that $\bm{R}^\prime(\bm{r}, \gamma)$ is isotropically distributed and spatially uncorrelated on the length scales of interest, as indicated by the microscopy experiments. Under these conditions, Eq.~\ref{eqn:IntensityCorrelation_Largeq} can be modified to take into account non-affinity by introducing a new Gaussian term:
%+++++++++++++++++++++++++++++++++++++++++++++++
\begin{equation}
    g_2(\bm{q}, \gamma) - 1 =\sinc^2\left(q_x \gamma \frac{e}{2}\right) \exp \left(-\frac{1}{3}q^2 c \gamma^2\right )\,.
    \label{eqn:DLS_nonAffineCorr}
\end{equation}
%+++++++++++++++++++++++++++++++++++++++++++++++
The argument of the exponential is motivated by the analogy between $c\gamma^2$, the non-affine msd under shear and $6Dt$, the msd in Brownian diffusion, for which~\cite{berne_dynamic_1976} $g_2-1 = e^{-2q^2Dt}$. The dashed lines in Fig.~\ref{fig:Affine_vs_nonaffine} show the correlation functions obtained for various $q$ via Eq.~\ref{eqn:DLS_nonAffineCorr}, with $c = 30 \um^2$. An excellent agreement is seen with both simulations (performed using the same $c$ value) and experiments, thus validating the modelling. The generalized diffusion coefficient $c$ found in DLS experiments is about one half of that measured by optical microscopy: since the two experiments are performed on distinct samples, this discrepancy most like stems from sample-to-sample variations in the spatial fluctuations of the elastic modulus.

Equation~\ref{eqn:DLS_nonAffineCorr} shows that the dominating contribution to the decay of $g_2-1$ depends on how fast each term on the r.h.s. decreases with increasing $\gamma$. Non-affinities are best measured when the Gaussian term decays faster than the $\sinc^2$ term, i.e. for $q_x < q \sqrt{c}/e$. In other words, DLS can detect non-affine rearrangements provided that the ROI used for the data analysis is `thin' enough, i.e. has a small enough size along the shear direction $\hat{u}_x$. Correlation functions obtained from `thick' ROIs, by contrast, will be dominated by the affine contribution.

\begin{figure}[h]
\centering
  \includegraphics[width=\columnwidth,clip]{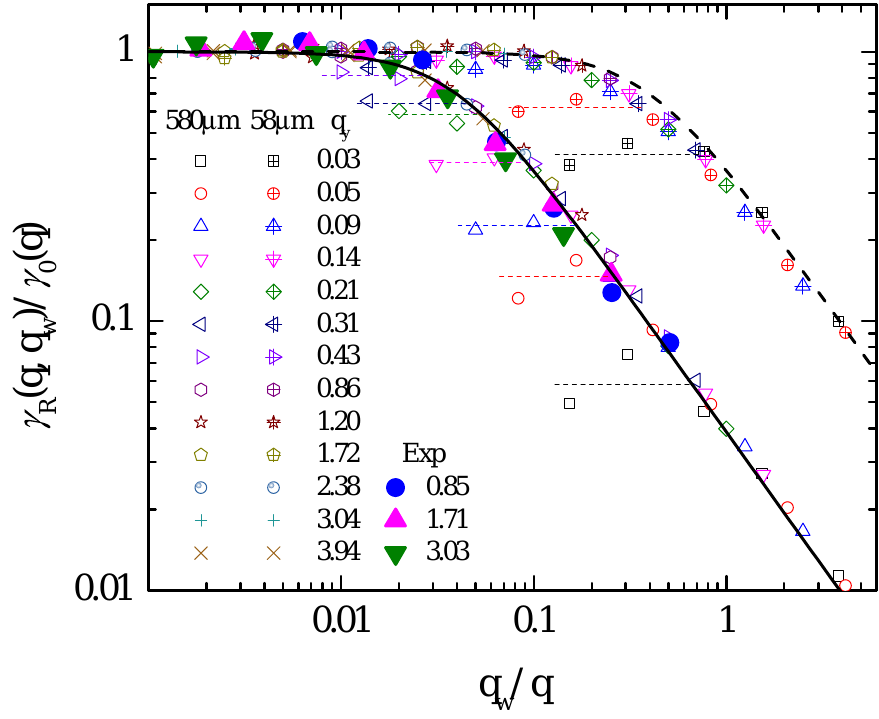}
  \caption{Characteristic strain for the decay of $g_2-1$ vs ROI width along the shear direction, for various $q_y$, as shown by the label (in $\um^{-1}$). For a given gap $e$, data collected at various $q$ collapse on a single curve when using scaled variables, as detailed in the text. The plateau region at low $q_w$ corresponds to the desired thin ROI regime where the relaxation of $g_2-1$ is controlled by non-affine displacements. Small symbols: numerical simulations, with $c=30\um^2$, $w = 450\um$, and $e=580 \um$ (open symbols) or $e= 58 \um$ (open symbols with cross). Large symbols: experimental data obtained by analyzing the correlation functions shown in Fig.~\ref{fig:Affine_vs_nonaffine} ($e=580 \um$, $c\approx 20\um^2$). Lines: theoretical curve, Eq.~\ref{eqn:DLS_nonAffineRescaling}, with $c = 30\um^2$ and $e=580 \um$ (solid line) or $e=58 \um$ (dashed line). }
  \label{fig:DLS_nonAffRescale}
\end{figure}

In practice, determining whether or not the thin ROI condition is met may not be trivial, since the upper bound for $q_x$ depends on $c$, which is not known \textit{a priori}. One way to address this issue is to quantify the characteristic shear deformation $\gamma_R$ at which $g_2-1$ decays and investigate how this quantity depends on the thickness of the ROI: the thin ROI limit will correspond to the regime where $\gamma_R$ is independent of the ROI thickness and the decay of $g_2-1$ is fully dominated by the second factor in the r.h.s. of Eq.~\ref{eqn:DLS_nonAffineCorr}.
We demonstrate this approach in Fig.~\ref{fig:DLS_nonAffRescale}, where we analyze the experimental and numerical correlation functions of
Fig.~\ref{fig:Affine_vs_nonaffine}. We quantify the thickness of the ROI by $q_w$, the maximum of the $q_x$ component associated to a ROI, as in the discussion of Fig.~\ref{fig:ROI_thickness}. We obtain $\gamma_R$ from a compressed exponential fit of the correlation function: $g_2(q, q_w,\gamma)-1 = \exp\left[-(\gamma/\gamma_R(q,q_w))^p\right]$, where we have explicitly indicated that $\gamma_R$ and thus $g_2-1$ depend on both the modulus of the scattering vector and the ROI thickness $q_w$. Figure~\ref{fig:DLS_nonAffRescale} shows $\gamma_R$ normalized by its $q_w \rightarrow 0$ limit, as a function of the the ROI width $q_w$ normalized by $q$. Two regimes are clearly seen for both simulations and experiments: at low $q_w/q$ (thin ROI regime), $\gamma_R$ is independent of $q_w$, while for thicker ROIs $\gamma_R$ decreases as $q_w^{-1}$, as expected when the $\sinc^2$ term in Eq.~\ref{eqn:DLS_nonAffineCorr} controls the decay of $g_2-1$.

By comparing the open and crossed small symbols obtained by simulating samples with two different gaps, one can see that the crossover between the two regimes depends on the sample thickness $e$, as expected from Eq.~\ref{eqn:DLS_nonAffineCorr}.
We find that an empirical expression that describes well the full behavior of $\gamma_R$ across the two regimes is given by
%+++++++++++++++++++++++++++++++++++++++++++++++
\begin{equation}
    \gamma_R^{-1}(\bm{q}, q_w)= \gamma_0^{-1}(\bm{q}) \sqrt{1+\left(\frac{q_w e}{5q\sqrt{c}}\right)^2} \,,
    \label{eqn:DLS_nonAffineRescaling}
\end{equation}
%+++++++++++++++++++++++++++++++++++++++++++++++
where $\gamma_0^{-1}=q\sqrt{c/3}$ is the characteristic strain expected for an infinitely thin ROI when finite beam size effects can be neglected (see Eq.~\ref{eqn:DLS_nonAffineCorr}), and the factor of 5 is introduced to obtain the best collapse with simulation data.  As shown by the solid and dashed lines in Fig.~\ref{fig:DLS_nonAffRescale}, this expression reproduces numerical and experimental data very well.

%The contribution of this faster decorrelation can be effectively observed for thin enough samples at scattering vectors with a predominant component perpendicular to the shear direction, i.e. when $q_x e < q \sqrt{c}$ and $q_y > \frac{e}{w\sqrt{c}}$ (we will comment later on on this second condition).

%That's why a practical way to quantitatively characterize nonaffinities is to look at the characteristic relaxation strain $\gamma_R$ extracted from the correlation functions for different ROI sizes: in principle it should scale as $q_w^{-1}$ for large ROIs ($q_w$ being the maximum $q_x$ vector contributing to the observed signal, as in figure \ref{fig:ROI_thickness}), and reach a $q_w$-independent plateau for ROIs thin enough. We propose as empirical analytic expression for the crossover the following:

%------------------------------------------------------------------- FIG
%------------------------------------------------------------------ FIG

So far, we have neglected the contribution due to the finite beam size, which was not included in Eq.~\ref{eqn:DLS_nonAffineCorr}. Indeed, even in the absence of non-affine rearrangements, a $q_w$-independent plateau would eventually be reached at low $q_w$ (thin ROI limit) when the translation $\delta$ of the moving plate becomes comparable to the beam size $w$, as discussed in Sec.~\ref{subsec:DLStrasl}. This corresponds to a critical strain $\gamma_{th}\approx  \frac{w}{e}$: for $\gamma \ge \gamma_{th}$, finite beam effects dominate and detecting any extra contribution due to non-affine displacements becomes increasingly difficult. Non-affine dynamics are therefore best seen when the characteristic strain $\gamma_0$ introduced in Eq.~\ref{eqn:DLS_nonAffineRescaling} is smaller than $\gamma_{th}$. Recalling that $\gamma_0=q^{-1}\sqrt{c/3}$ and that $q \approx q_y$ in experiments designed to detect non-affine displacements, the condition $\gamma_0 \le \gamma_{th}$ is re-casted in the form $q_y \gtrsim e/(w\sqrt{c})$ (where we have dropped a factor of $\sqrt{3}$ for simplicity), which may be fulfilled by decreasing the gap and increasing the beam size.

The effect of the finite beam size can be seen in the simulation data of Fig.~\ref{fig:DLS_nonAffRescale} for the smallest $q$ vectors (see e.g. data for $q_y \le 0.43\um^{-1}$ for the gap $e = 580\um$). As shown by the horizontal dashed lines of Fig.~\ref{fig:DLS_nonAffRescale}, for these scattering vectors, $\gamma_R$ deviates from the master curve described by Eq.~\ref{eqn:DLS_nonAffineRescaling} and saturates at values much lower than $\gamma_0(q)$, the expected $q_w \rightarrow 0$ limit. Note that for the thinner cell with $e = 58\um$, the effect of the finite beam size becomes relevant only at much lower $q$ vectors (see the crossed data points for $q_y \le 0.05\um^{-1}$), in agreement with the scaling with the gap of the $q_y \gtrsim e/(w\sqrt{c})$ condition.

To recapitulate the main findings of this section, we have shown that non-affine displacements may be resolved by measuring correlation functions for $q$ vectors oriented perpendicular to the shear direction. Non-affine dynamics can be unambiguously quantified by DLS provided that i) the width of the ROI in the shear direction is small enough ($q_w  < q \sqrt{c}/e$); ii) the $y$ component of the probed scattering vector is large enough or the sample thin enough ($q_y \gtrsim e/(w\sqrt{c})$). These bounds depend on the generalized diffusion coefficient $c$, which is not known a priori. However, the scaling plot of Fig.~\ref{fig:DLS_nonAffRescale} and Eq.~\ref{eqn:DLS_nonAffineRescaling} provide consistency checks allowing one to verify a posteriori whether or not the measured dynamics is due only to non-affine displacements.

\section{Conclusions}
\label{sec:conclusions}

In this work we have shown how DLS may be coupled to a shear cell in order to probe affine displacements and non-affine rearrangements. Correlation functions measured for a $q$ vector oriented parallel to the direction of the applied shear are in principle sensitive to both affine and non-affine displacements. In practice, however, affine displacements typically dominate over non-affine ones (at least in the interesting regime corresponding to the onset of non-linear behavior), such that data for $q \approx q_x$ essentially probe affine displacements. Measuring the affine component is a powerful tool to check for the occurrence of wall slip or shear banding: any deviation from an ideal affine deformation profile would result in correlation functions that depart from the expected theoretical form discussed in the previous sections.

Non-affine dynamics are accessible by measuring correlation functions for a $q$ vector oriented perpendicularly to the shear direction. Under realistic conditions, these correlation functions may contain additional contributions not due to non-affine displacements. These unwanted contributions can be neglected under appropriate experimental conditions. The most important role is played by the finite $q_x$ component of the ROIs over which $g_2-1$ is averaged. This component has to be minimized in order to suppress the spurious decay due to the affine deformation. In practice, choosing ROIs with a width along $\hat{u}_x$ smaller than about 5 pixels is sufficient. We emphasize that the relevant parameter is the absolute value of the ROI width, not the ratio $q_x/q_y$. Thus, the same (small) width should be used for all ROIs, irrespective of the magnitude of the scattering vector.

Another source of spurious decorrelation stems from the rigid translation of scatterers. A cell with counter-moving plates and a stagnation plane in $z=e/2$ would reduce this effect. Furthermore, this contribution may be reduced by increasing the lateral size of the beam and by decreasing the sample thickness. Thinner samples also help in mitigating the effect of the finite ROI width; additionally, in the $e \ll w$ regime the decay of $g_2-1$ is easier to describe with analytical models. Finally, optical aberrations and the curvature of the wavefront impinging on the sample spuriously accelerate the decay of the correlation function. This artifact can be reduced by placing the sample in the beam waist, where the wavefront is plane, and can be fully corrected for using the spatio-temporal crosscorrelation method described in Sec.~\ref{subsec:DLStrasl}. Similarly to the term arising from the sample translation, this contribution becomes negligible in the thin sample limit, $e\ll w$.

As a final remark, we note that although some care must be taken in order to optimize the experimental parameters, none of the required conditions  discussed above is impractical to meet. Indeed, typical small-angle light scattering apparatuses use a beam with $w$ in the range 0.5 - 10 mm. When coupled to a shear cell with a gap $e$ in the range 0.3-1 mm, which can be easily achieved, such an apparatus will operate in the thin sample limit, where all the above conditions are easily met for the typical range of accessible $q$ vectors. We thus hope that DLS coupled to rheology will become an increasigly popular method to probe the microscopic dynamics of soft systems under shear.

\section*{Acknowledgments}
We thank T. Phou for help in preparing the PA gels. This work was funded by ANR (grant n. ANR-14-CE32-0005-01), CNES, and the EU (Marie Sklodowska-Curie ITN Supolen, Grant No.
607937).

\section*{Appendix 1: list of symbols}
%\textbf{XXX Stefano: riordinare la lista in ordine alfabetico (prima alfabeto latino, poi alfabeto greco) XXX}
\begin{tabular}{ l l }
%\begin{longtable}{ l l }
	$\hat{u}_x, \hat{u}_y, \hat{u}_z$ & unit vectors in the shear, vorticity and shear \\
               			& gradient directions, respectively\\
	$\bm{r}$		& coordinates in the scattering volume \\
	$\Delta\bm{r}(\bm{r})$ & particle displacement field \\
	$\bm{R}'(\bm{r})$	& non-affine displacement field \\
	$\gamma$ 		& macroscopic shear deformation \\
	$e$ 			& sample thickness  \\
	$k$ 			& laser wave vector \\
	$\bm{q}$ 		& scattering vector\\
	$q_x, q_y, q_z$	& scattering vector components along $\hat{u}_x, \hat{u}_y, \hat{u}_z$ \\
	$\bm{E}_{in}$	& (complex) incident field \\
	$\bm{E}_S$		& (complex) scattered field \\
	$a$			& Rayleigh range \\
	$w_0$ 		& $1/e^2$ beam radius in the beam waist \\
	$w(z)$		& $1/e^2$ beam radius at distance $z$ from the waist \\
	$w$ 			& beam radius on the sample plane \\
	$\rho(z)$		& radius of curvature of the wavefront at position $z$\\
	$\bm{\delta}$	& translation of a 2D sample along the $\hat{u}_x$ direction \\
	$\bm{X}$		& relative position of two points on the sensor for\\
				& spatio-temporal correlation function (Eq.~\ref{eqn:CrossCorr_ref})\\
	$\bm{X}_{peak}$	& position of the spatio-temporal crosscorrelation peak \\
	$\Delta$ 		& speckle size\\
	$l$			& modified spatial correlation length (Eq.~\ref{eqn:ModifiedCorrelationLength_Article}) \\
	$\varepsilon$ 	& distance between the detector and the focal plane,\\
                & see Fig.~\ref{fig:setup}b)\\
	$R$			& distance between the detector and the sample\\
				& image, see Fig.~\ref{fig:setup}b)\\
	$\sigma$ 		& $=1+R/\rho$ \\
	$\bar{G}$			& spatially averaged shear modulus \\
	$G(\bm{x})$		& local shear modulus \\
	$\delta G(\bm{x})$ & local shear modulus fluctuations \\
	$\mathcal{A}$ 	& non-affine parameter (non-affine MSD) \\
	$c$			& generalized diffusion coefficient (non-affine\\
                			& MSD per unit squared strain, Eq.~\ref{eqn:NonaffineParameter})\\
	$D$    &   Brownian diffusion coefficient \\
	$q_w$		& detector size (in $q$ space) along $\hat{u}_x$ \\
	$\gamma_R(\bm{q}, q_w)$ & deformation needed to decorrelate the signal\\
                			& in presence of nonaffinities (Eq.~\ref{eqn:DLS_nonAffineRescaling})\\
	$\gamma_0(\bm{q})$ & thin ROI limit of $\gamma_R$ when finite beam size effects\\
				& can be neglected \\
	$\gamma_{th}$ 	& deformation needed to decorrelate the signal in\\
				& the thin ROI limit, for a pure affine deformation \\
\end{tabular}

\section*{Appendix 2: $g_2-1$ for an affine deformation}
\label{sec:nastyeq}

Equation~\ref{eqn:IntensityCorrelation_Largeq} of the main text can be easily computed starting from the field correlation function (Eq.~\ref{eqn:FieldCorrelation_def}) evaluated by using the scattered field of Eq.~\ref{eqn:scattered} and the affine particle displacement of Eq. \ref{eqn:AffineDeformation}. We start by assuming an ideal plane wave as the incident electric field, $\bm{E_{in}}(\bm{r})=\bm{E}_0$. The crucial step in the derivation is to realize that the double sum $\sum_{j,l}$ over the scatterers resulting from injecting Eq.~\ref{eqn:scattered} into Eq.~\ref{eqn:FieldCorrelation_def} actually reduces to a single sum over the self terms ($i=l$), whereas the cross terms ($i\neq l$) vanish for uncorrelated scatterer positions. This simple sum can be evaluated by casting it into an integral weighted by the density functional: $n(\bm{r})=\sum_j\delta_3(\bm{r}-\bm{r}_j)$, where $\delta_3(\bm{r})$ is the three-dimensional Dirac's delta. If the scatterers are distributed homogeneously in the scattering volume, the integral reduces to:
%+++++++++++++++++++++++++++++++++++++++++++++++
\begin{equation}
     g_1(\bm{q}, \gamma) = \frac{1}{e} \int\limits_{-e/2}^{e/2} e^{i\gamma q_x z'} dz' = \sinc\left(\gamma q_x \frac{e}{2}\right)\,,
    \label{eqn:FieldCorrelationShear_Integral}
\end{equation}
%+++++++++++++++++++++++++++++++++++++++++++++++
which is Eq.~\ref{eqn:IntensityCorrelation_Largeq} of the main text.

A correction for the finite beam size may be obtained by replacing the plane wave with a more realistic Gaussian profile (Eq.~\ref{eqn:GaussianBeam}), which unfortunately results in a rather involved expression. Progress can be made by assuming that the sample lays in the beam waist ($z=0$, $\rho=\infty$). In this case, Eq.~\ref{eqn:FieldCorrelationShear_Integral} becomes
%+++++++++++++++++++++++++++++++++++++++++++++++
\begin{equation}
     %g_1(\bm{q}, \gamma) \propto  e^{-2q_x^2w_0^2\frac{\left(\frac{\gamma}{2}\right)^2}{1+\left(\frac{\gamma}{2}\right)^2}}
     %\int\limits_{-\frac{e}{2}}^{\frac{e}{2}} dz' e^{-\frac{2}{w_0^2}\left[z'+iq_x w_0^2\frac{\frac{\gamma}{2}}{1+\left(\frac{\gamma}{2}\right)^2}\right]^2 \left[1+\left(\frac{\gamma}{2}\right)^2\right]} \,.
     g_1(\bm{q}, \gamma) \propto \int\limits_{-\frac{e}{2}}^{\frac{e}{2}} dz' e^{i q_x \gamma z'} \exp\left \{-2  \left[1+\left(\frac{\gamma}{2}\right)^2\right] \left(\frac{z'}{w_0}\right)^2\right \} \,.
    \label{eqn:FieldCorrelationShear}
\end{equation}
%+++++++++++++++++++++++++++++++++++++++++++++++
Here, the proportional sign indicates that the expression has to be properly normalized, such that $g_1(\bm{q}, 0)=1$.
Equation \ref{eqn:IntensityCorrelation_Largeq} corresponds to the limits $q_x\gg \frac{e}{\gamma w_0^2}$ and $e\ll w_0$. The opposite limit, $q_x\rightarrow 0$, is the desired condition for probing non-affine displacements. In this limit one has
%+++++++++++++++++++++++++++++++++++++++++++++++
\begin{equation}
     g_1(\bm{q}, \gamma) = \frac{\mathrm{Erf}\left[\frac{e}{w_0\sqrt{2}}\sqrt{1+\left(\frac{\gamma}{2}\right)^2}\right]}{\sqrt{1+ \left(\frac{\gamma}{2}\right)^2} \mathrm{Erf}\left(\frac{e}{w_0\sqrt{2}}\right)}
    \label{eqn:FieldCorrelationShearQperp}
\end{equation}
%+++++++++++++++++++++++++++++++++++++++++++++++
For thin samples ($e \lesssim w_0$) this expression decays when $\gamma \frac{e}{2} \gtrsim w_0$, which corresponds to the requirement that the absolute displacements are comparable to the beam size. However, one should be aware that this limit cannot be fully reached in practice, since  $q_x$ is bounded from below by the speckle size, and thus cannot be smaller than $q_{min}\sim w_0^{-1}$. By taking this limiting value, the validity of Eq.~\ref{eqn:FieldCorrelationShearQperp} is restricted to $\gamma \ll \frac{e}{w_0}$. While these approximated forms are useful to rapidly grasp the general behavior of $g_2-1$, the general form, Eq. \ref{eqn:FieldCorrelationShear}, can be easily be integrated numerically to obtain precise theoretical predictions, as shown in the main text.

\footnotesize{
%\bibliography{biblio_DLS}
%\bibliography{drops}
%your .bib file
%\bibliographystyle{rsc}
%the RSC's .bst file
\providecommand*{\mcitethebibliography}{\thebibliography}
\csname @ifundefined\endcsname{endmcitethebibliography}
{\let\endmcitethebibliography\endthebibliography}{}

}

\end{document}